\documentclass[11pt,iop]{emulateapj}
\usepackage{amsmath}
\usepackage{ulem}
\usepackage{graphicx}
\usepackage{natbib}
\begin{document}

\title{Constraints on the Photon Mass with Fast Radio Bursts}
\author{Xue-Feng Wu$^{1,2}$, Song-Bo Zhang$^1$, He Gao$^3$, Jun-Jie Wei$^1$,
Yuan-Chuan Zou$^4$, Wei-Hua Lei$^4$, Bing Zhang$^5$, Zi-Gao Dai$^6$, and Peter M{\'e}sz{\'a}ros$^{7,8,9}$}

\altaffiliation{$^1$ Purple Mountain Observatory, Chinese Academy of Sciences, Nanjing 210008, China; xfwu@pmo.ac.cn\\
$^2$ Joint Center for Particle, Nuclear Physics and Cosmology, Nanjing University-Purple Mountain Observatory, Nanjing 210008, China \\
$^3$Department of Astronomy, Beijing Normal University, Beijing 100875, China \\
$^4$ School of Physics, Huazhong University of Science and Technology, Wuhan 430074, China \\
$^5$ Department of Physics and Astronomy, University of Nevada Las Vegas, Las Vegas, NV 89154, USA \\
$^6$ School of Astronomy and Space Science, Nanjing University, Nanjing 210093, China \\
{$^{7}$}Department of Astronomy and Astrophysics, Pennsylvania State University, 525 Davey Laboratory, University Park, PA 16802 \\
{$^{8}$}Department of Physics, Pennsylvania State University, 104 Davey Laboratory, University Park, PA 16802 \\
{$^{9}$}Center for Particle and Gravitational Astrophysics, Institute for Gravitation and the Cosmos, Pennsylvania State University, 525 Davey
Laboratory, University Park, PA 16802 \\
}


\begin{abstract}
Fast radio bursts (FRBs) are radio bursts characterized by millisecond durations,
high Galactic latitude positions, and high dispersion measures.
Very recently, the cosmological origin of FRB 150418 has been confirmed by \cite{kea16},
and FRBs are now strong competitors as cosmological probes. The simple sharp
feature of the FRB signal is ideal for them to probe some of the fundamental laws of physics.
Here we show that by analyzing the
delay time between different frequencies, the FRB data can place stringent upper limits on the rest mass
of the photon. For FRB 150418 at $z=0.492$,
one can potentially reach $m_{\gamma} \leq 5.2 \times 10^{-47}$ g, which is $10^{20}$ times smaller than the rest mass of
electron, and is about $10^3$ times
smaller than that obtained using other astrophysical sources with the same method.
\end{abstract}


\section{Introduction} \label{sec1}
Maxwell's equations, the successful theory of classical electromagnetic mechanical, have a fundamental prediction that all electromagnetic radiation propagate in vacuum at the constant speed $c$, independent of frequency of the wave. This is also adopted as the second postulate of Einstein's theory of special relativity. As a result, the rest mass of the photon, the fundamental quanta of electromagnetic fields, should be strictly zero.

Testing the correctness of this prediction is one of the most intriguing tasks
in modern physics. It is closely related to many fundamental questions such as charge conservation
and quantization, the possibility of charged black holes, etc. \citep{goldhaber1971terrestrial, 2005RPPh...68...77T}, and hence, the need to
push its verification as far as possible.  According to the uncertainty principle, it is impossible to do
any experiment that would firmly establish that the photon rest mass is exactly zero. The ultimate upper
limit on the photon rest mass would be $m_{\gamma} \leq \hbar/(\Delta t)c^2\approx10^{-66}~\rm{g}$, when the age
of the Universe ($\sim 10^{10}$ years) is used.

From the theoretical point of view, a nonzero photon mass could be accommodated in a unique way by changing
the inhomogeneous Maxwell's equations to the Proca equations \citep{proca1936theorie, 1984Natur.307...14B}. Gauge invariance is
replaced by the Lorentz gauge, so that a mass term can be added to the Lagrangian density for the
electromagnetic field by invoking a characteristic length associated with the photon rest mass,
$\mu_{\gamma}^{-1}=\hbar/m_{\gamma}c$, to describe the effective range of the electromagnetic
interaction \citep{proca1936theorie}. In this case, the electric and magnetic potentials themselves have a physical
significance, not just through their derivatives, which would lead to far-reaching implications. For
instance, the speed of light in free space would depend on its frequency, longitudinal electromagnetic waves
could exist, magnetic dipole fields would suffer more rapid fall-off with distance due to the addition of a
Yukawa component to the magnetic potential, and so on. All of these effects have been employed to derive
increasingly restrictive constraints on the photon rest mass, either through laboratory experiments or
astrophysical and cosmological observations \citep{goldhaber1971terrestrial, 2005RPPh...68...77T, 2012PhRvL.109m1102P}.

The most direct method of constraining the photon mass is to detect a possible frequency dependence of the
speed of light.
When the laboratory conditions eventually impose a limit,
astronomical events afford the best opportunity for obtaining higher precision measurements on the relative
speed of electromagnetic radiation at different wavelengths. The first attempt that took advantage of
astronomical distances was the comparison of the arrival time of optical and radio emission from flare
stars \citep{1964Natur.202..377L}, with a constraint on the photon mass of $m_{\gamma}\leq1.6\times10^{-42}$ g. With a
measurement of the dispersion in the arrival time of optical wavelengths of 0.35 and 0.55 $\rm \mu m$ from
the Crab Nebula pulsar, a stringent limit on the possible speed dependence on frequency was set \citep{1969Natur.222..157W},
but the corresponding limit on the photon mass was only $m_{\gamma}\leq5.2\times10^{-41}$ g. Assuming that
the radio and gamma-ray photons of gamma-ray bursts (GRBs) have the same origin and that they are emitted at
the same time, \cite{1999PhRvL..82.4964S} set the most stringent limit on the frequency dependence of speed of light
up to now, implying a photon mass $\leq 4.2\times10^{-44}$ g, by analysing the arrival time delay between
radio and the gamma-ray emission from GRB 980703 at high redshift.

The discovery of Fast Radio Bursts (FRBs) \citep{2007Sci...318..777L, 2013Sci...341...53T} makes it possible to apply these new probes to constrain
the photon mass. The pulse arrival times follow the $\nu^{-2}$ law, which is consistent with the propagation of radio
waves through a cold plasma. The high values of DM (dispersion measure), if contributed from
the inter-galactic medium (IGM), would place these bursts in the redshift range of 0.5 to 1 \citep{2013Sci...341...53T}. 
Even though the distance scale and physical origin of these events are still subject to debate,
the extra-galactic origin of FRBs is suggested by an increasing number of recent observations \citep{2014ApJ...797...70K, kulkarni2015arecibo, masui2015dense, margalit2015inferring, saint2014short},
especially the detection of ten additional bursts from the direction of FRB 121102 \citep{2016arXiv160300581S},
and the potential applications of FRBs as cosmological probes have been suggested \citep{2014ApJ...783L..35D, 2014ApJ...788..189G, 2014ApJ...780L..33M, zheng2014probing,zhou2014fast,2015MNRAS.451.3278M}.
Very recently, \cite{kea16} report the first precise localization for an FRB thanks to the
identification of a fading radio transient, providing the first redshift measurement to FRB 150418 ($z=0.492\pm0.08$).\footnote{
However, \cite{2016arXiv160208434W} suggest that the proposed radio transient is a common Active Galactic
Nuclei (AGN) variability and unrelated to FRB 150418, and hence that the redshift determination may not be justified.
While, it has been proposed that the chance probabilities to have such a highly variable AGN in the Parkes beam and to have the radio source just start to decay after the FRB are low \citep{2016arXiv1603.04825L}. Also the early fading of the radio transient can be understood as an afterglow of a relativistic explosion with energy comparable to a short GRB \citep{2016arXiv1602.08086Z}.}
The simple, sharp, temporal feature of the FRB signal allows one to easily derive the observed time delay
between different frequencies. These time lags are
usually small ($\sim 1$ s), which make FRBs one of the best candidates for
constraining the rest mass of photon.

In this letter, we study the prospects of constraining the photon rest mass with FRBs in detail.
Our original motivation is the discovery of a radio wave in FRB 150418 with an approximately
0.815 s delay time that occurs simultaneously from frequencies 1.2 to 1.5 GHz.
We will show that FRBs can be a new powerful tool for constraining the photon mass.
Throughout this letter, we adopt the cosmological parameters recently derived
from nine-year $Wilkinson$ $Microwave$ $Anisotropy$ $Probe$ (WMAP) observations \citep{2013ApJS..208...19H}:
$\Omega_{\rm m}=0.286$, $\Omega_{\Lambda}=0.714$, $\Omega_{\rm b}=0.046$, and $h_{0}=0.69$.

\section{Velocity dispersion by the non-zero mass of photon} \label{sec2}
Assuming that the photon has a non-zero rest mass $m_{\gamma}$, according to Einstein's
special relativity, the energy equation can be written as
\begin{equation}\label{eq1}
E=h\nu=\sqrt{p^2c^2+m_\gamma^2c^4}\;.
\end{equation}
In vacuum, the dispersion relation between the speed of photon $\upsilon$ and frequency $\nu$ is
\begin{equation}\label{eq2}
\upsilon=\frac{\partial{E}}{\partial{p}}=c\sqrt{1-\frac{m_\gamma^2c^4}{E^2}}\approx c\left(1-\frac{1}{2}A\nu^{-2}\right)\;,
\end{equation}
where $A=m_\gamma^2c^4/h^2$.
Equation~(\ref{eq2}) shows that the high energy/frequency photons travel in vacuum faster than the low energy/frequency photons.
Two photons emitted simultaneously by a source would arrive on Earth with a time delay if they have different frequencies.

Here we consider two photons (one with higher frequency $\nu_{h}$ and the other
with lower frequency $\nu_{l}$) are emitted simultaneously
from a source at redshift $z$. For the high energy photon, the time of arrival to the Earth is set to be $z=0$.
Since the higher energy photon would arrive earlier than the lower one, the arrival time corresponds to
$z=-\Delta z$ $(\Delta z \ll 1)$ for the lower one. Taking into account of the
cosmological expansion, one can derive the comoving distance from the source to the Earth
$
  x(z,\nu_h)=\frac{c}{H_0}
  \int_0^{z}\left[1-\frac{1}{2}A\nu_h^{-2}\frac{1}{(1+z')^2}\right]
  \frac{d{z'}}{\sqrt{\Omega_{\rm m}(1+z')^3+\Omega_{\Lambda}}}
$
for the higher energy photon, and
$
  x(z,\nu_l)=\frac{c}{H_0}
  \int_{-\Delta{z}}^{z}\left[1-\frac{1}{2}A\nu_l^{-2}\frac{1}{(1+z')^2}\right]
  \frac{d{z'}}{\sqrt{\Omega_{\rm m}(1+z')^3+\Omega_{\Lambda}}}
$
for the lower energy photon.
Because the comoving distance traveled by these two photons should be the same (i.e., $x(z,\nu_h)=x(z,\nu_l)$),
the observer-frame delay time ($\Delta{t_{m_{\gamma}\neq 0}}$) between these two photons can be expressed as
\begin{equation}\label{eq5}
  \Delta{t_{m_{\gamma}}}=\frac{\Delta{z}}{H_0}=
  \frac{A}{2H_0}\left(\nu_l^{-2}-\nu_h^{-2}\right)H_1(z)\;,
\end{equation}
where $H_0=100 h_{0}$ km $\rm s^{-1}$ $\rm Mpc^{-1}$ is the Hubble constant and
\begin{equation}\label{}
H_1(z)=\int_0^{z}\frac{(1+z')^{-2}dz'}{\sqrt{\Omega_{\rm m}(1+z')^3+\Omega_{\Lambda}}}\;.
\end{equation}
So the photon mass can be constrained as
\begin{equation}\label{}
 m_{\gamma}=hc^{-2}\left[\frac{2H_0 \Delta{t_{m_{\gamma}}}}{\left(\nu_l^{-2}-\nu_h^{-2}\right)H_1(z)}\right]^{1/2}\;,
\end{equation}
which can be simplified as
\begin{equation}\label{eqmgamma1}
m_{\gamma}=\left(1.56\times10^{-47}\rm g\right)
\left[\frac{\Delta{t_{m_{\gamma}}}}{\left(\nu_{l,9}^{-2}-\nu_{h,9}^{-2}\right)H_1(z)}\right]^{1/2}\;,
\end{equation}
where $\nu_{9}$ is the radio frequency in units of $10^{9}$ Hz.

\begin{figure}[h]
\vskip-0.2in
\centerline{\includegraphics[angle=0,scale=0.6]{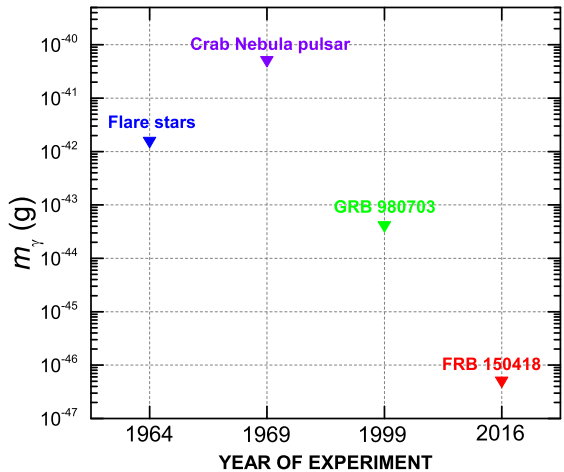}}
\vskip-0.1in
\caption{Strict upper limits on the photon rest mass from detection of the frequency dependence of
speed of light from astrophysical sources. The upper limits from flare stars, Crab nebula pulsar and GRB 980703 are taken from
\cite{1964Natur.202..377L}, \cite{1969Natur.222..157W}, and \cite{1999PhRvL..82.4964S}, respectively.}\label{f1}
\end{figure}

FRB 150418 was detected by the Parkes radio telescope on 2015 April 18
UTC \citep{kea16}. Rapid multi-wavelength follow-up identified a fading
radio afterglow, which was used to identify the host galaxy.  At the position of the afterglow,
imaging on the Subaru and Palomar 200-inch telescope and spectroscopy
with FOCAS on Subaru and DEIMOS on Keck reveal the galaxy's redshift to be $z=0.492$.
From the frequency-dependent delay of FRB 150418 (see Figure~1 of Ref.~\cite{kea16}),
we can obtain an approximate $\Delta t=0.815$ s delay time between frequencies $\nu_l = 1.2$ GHz and $\nu_h = 1.5$ GHz.
In principle, the total delay time may contain several contributions \citep{2015ApJ...810..121G, 2015PhRvL.115z1101W}, including, e.g.,
an intrinsic (astrophysical) time delay between two test photons, a time delay caused by effects of
Lorentz invariance violation (if this exists), and a time delay from the dispersion process
by the line-of-sight free electron content. Assuming that all of the observed time lag can be attributed to a
non-zero photon mass, a conservative upper limit for $m_{\gamma}$ can be estimated.\footnote{
With the similar approach, i.e., based on knowledge of the FRB redshift
and assuming the time delay is dominated by the gravitational field of the Milky Way,
\cite{2016ApJ...820L..31T} set a strict limit on the Einstein Equivalence Principle from FRB 150418.}
As shown in Fig. \ref{f1}, a strict limit on the photon rest mass from FRB 150418 is
$m_{\gamma} < 5.2\times10^{-47}$ g, which is $10^{3}$ times better
than that obtained by GRBs \citep{1999PhRvL..82.4964S}.
We note that our restriction has been generally confirmed later by \cite{2016arXiv160209135B}.

\section{Velocity dispersion by the plasma effect} \label{sec3}

According to the
characteristic observational feature of FRBs, i.e., that the arrival time delay at a given frequency
$\nu$ follows a $\nu^{-2}$ law, the observed time delay $\Delta t$ between different frequencies should be mainly
contributed by the non-zero photon mass effect and the plasma effect via the dispersion process
from the line-of-sight free electron content, i.e., $\Delta t \approx \Delta t_{m_{\gamma}} + \Delta t_{\rm plasma}$. Note that $\Delta t_{\rm plasma}$ has the same sign as  $\Delta t_{m_{\gamma}}$, as both predict higher energy photons travel faster than the lower energy photons (see below).
In our calculation, we have for simplicity assumed that all
of the observed time delay is caused by the non-zero photon mass effect.
For completeness, we will also test the case by subtracting the contribution from the plasma effect in $\Delta t$.
It should be underlined that the measured dispersion has contributions from the Milky Way, intervening intergalactic medium (IGM),
and FRB host galaxy. Since the number density of electrons in host galaxy is hard to know and the contribution of the Milky Way
dispersion is negligible, we just consider the IGM dispersion.

We assume that the intervening intergalactic medium between
the source and the Earth is fully ionized hydrogen and
singly-ionized Helium when the source located very late after
the re-ionization epoch of the Universe (e.g., $z<3$), and
the ionized IGM is homogeneously distributed. The plasma frequency of
IGM is $\omega_{p}=\sqrt{4\pi e^2 n_e/m_e}=5.64\times10^4
n_e^{1/2}$ s$^{-1}$. The number density of electrons in IGM
increases with the redshift, $n_e=n_e(0)(1+z)^3$. We adopt
the mass fraction of Helium in IGM is $Y=0.24$ \citep{2003ApJ...598...73Y},
the free electron number per baryon can be estimated to be 0.82.
So the local number density of free electrons in IGM is
$n_e(0)=0.82n_b(0)=9.23\times10^{-6}\Omega_{b}h_{0}^2$ cm$^{-3}$ \citep{2004MNRAS.348..999I},
where $n_b(0)=3\Omega_{b}H_0^2/(8\pi Gm_p)$ is the local baryon number density
and $\Omega_b$ is the current baryon fraction of the Universe.
The IGM magnetic effect on the velocity dispersion of photons can be neglected, since the Lamour
frequency $\omega_{L}=eB/m_e c=17.6(B/\mu{\rm G})$ s$^{-1}$ is
much smaller than $\omega_{p}$ for typical IGM with
$n_{e}>10^{-6}$ cm$^{-3}$ and $B<10^{-6}$ G. Due to the plasma
effect, the speed of a photon with energy $E$ travelling in IGM at redshift $z$
is
\begin{equation}\label{eq7}
\upsilon_p(E)=\displaystyle
c\sqrt{1-\left(\displaystyle\frac{E_p(z)}{E(z)}\right)^2}
\cong\left[1-\frac{\nu_p^{2}(0)(1+z)}{2\nu^{2}}\right],
\end{equation}
for $E(z)>E_{p}(z)$ (i.e., $\nu(z)>\nu_{p}(z)$), where $\nu_{p}(z)=(1+z)^{3/2}\nu_{p}(0)$, $\nu(z)=(1+z)\nu(0)$, and
\begin{equation}
\nu_p^{2}(0)=\left(\frac{\omega_p(0)}{2\pi}\right)^{2}=\frac{0.82e^{2}n_b(0)}{{\pi}m_e}\;.
\end{equation}
This dispersion also leads photons with higher energies travelling
faster than those with lower energies (see Equation~(\ref{eq7})).

Similar to the derivations of the formulas in Section 2,
the time lag caused by the IGM plasma effect can be expressed as
\begin{equation}\label{eq15}
\Delta t_{\rm plasma}=\displaystyle\frac{\Delta
z}{H_0}=\frac{\nu_p^{2}(0)}{2H_0}\left(\nu_l^{-2}-\nu_h^{-2}\right)H_2(z)\;,
\end{equation}
where
\begin{equation}
H_2(z)=\int_{0}^{z}\frac{(1+z')dz'}{\sqrt{\Omega_{\rm m}
(1+z')^3+\Omega_{\Lambda}}}\;.
\end{equation}
Equation~(\ref{eq15}) can be simplified as
\begin{equation}
\Delta t_{\rm plasma}\approx3.64\;{\rm s}\left(\frac{\Omega_bh_{0}}{0.032}\right)H_2(z)\left(\nu_{l,9}^{-2}-\nu_{h,9}^{-2}\right).
\end{equation}

With the redshift of FRB 150418, an estimation on $\Delta t_{\rm plasma}$ between frequencies $\nu_l = 1.2$ GHz and $\nu_h = 1.5$ GHz from Equation~(11)
is $\Delta t_{\rm plasma}=0.491$ s. Now the stricter upper limit on the photon mass comes to be
$m_\gamma\leq3.3\times10^{-47}$ g for the case of $\Delta t_{m_{\gamma}}=\Delta t-\Delta t_{\rm plasma}=0.324$ s.
If we have a better understanding of the host galaxy dispersion, our constraint would be further improved in some degree.

\section{Summary and Discussion} \label{sec5}
In this letter, we have shown that FRBs can be used to
place severe limits on the photon mass $m_\gamma$.
From the frequency-dependent
delay of FRB 150418, we obtained approximately a delay time $(\Delta{t})=0.815$ s
of the radio pulse at $\nu_l=1.2$ GHz relative to that at $\nu_h=1.5$ GHz.
Considering the delay time was caused by the non-zero photon mass ($m_{\gamma}\neq0$) effect
and adopting the possible redshift $z=0.492$ for FRB 150418, the severe limit on the photon mass is
$m_{\gamma} \leq 5.2 \times10^{-47}$ g, which represents an improvement of three
orders of magnitude over the results by GRBs from \cite{1999PhRvL..82.4964S}.

Notice that this is a conservative upper limits: the inclusion of contributions from the other neglected
potential contributions to $\Delta t$ can make this limit even more stringent. If the time delay between different frequencies is mainly contributed by the plasma effect,
more severe constraints would be achieved (e.g., $m_\gamma\leq3.3\times10^{-47}$ g)
if the contribution from the plasma effect has been subtracted in $\Delta t$ (i.e., $\Delta t_{m_{\gamma}}=\Delta t-\Delta t_{\rm plasma}$).
Moreover, \cite{kea16} found that the measurement of the cosmic density of ionized baryons
$\Omega_{\rm IGM}$ from the dispersion measure and the redshift of FRB 150418 is in agreement with the expectation from
WMAP observations, which leads one to conclude that
the observed time delay for FRB 150418 is highly dominated by plasma
dispersion, with very small contribution possibly from the non-zero photon mass effect.
Optimistically, if the non-zero photon mass effect is responsible for $10.0\%$ of $\Delta t$ of FRB 150418,
the upper limit on $m_{\gamma}$ could be set to $m_{\gamma}< 1.6\times10^{-47}$ g, which is closer to the current most stringent constraints as adopted by the
Particle Data Group \citep{amsler2008review}. The results presented here show the potentially high
benefits to be obtained when more FRBs are observed, especially if the redshifts of FRBs can be more likely to be measured
in the future.

While there may be multiple physical origins for the population of FRBs,
the extra-galactic origin of at least some FRBs is receiving increasingly supports from the
observational data \citep[e.g.,][]{2016arXiv160300581S}. We find that even if FRB 150418
originate within our Local Group \citep[1 Mpc;][]{2006Ap.....49....3K}
or the Local Supercluster \citep[50 Mpc;][]{2014Natur.513...71T},
a strict limit on the photon mass of $m_{\gamma} < 1.9\times10^{-45}$ g or $m_{\gamma} < 2.7\times10^{-46}$ g
can be still achieved, which is already 10 times or 100 times
smaller than that obtained by GRBs using the same method \citep{1999PhRvL..82.4964S}.

Currently, the detection rate of FRBs is relatively low, mainly
due to the lack of either necessary high time resolution or a wide field of view in any of the current
telescopes. Future radio transient surveys such as the Square Kilometer Array would break through these
obstacles \citep{dewdney2009square}, and are expected to discover and precisely localize an increasing number of FRBs.
With the abundant observational information in the future, the mysteries of the physical nature of FRBs
are expected to be eventually unveiled, and their capability for testing fundamental physics, as discussed
here, will find increased use.

\acknowledgments
We are grateful to the anonymous referees for insightful comments and helpful suggestions to improve the paper.
We thank Qingwen Wu and Biping Gong for their helpful discussion. This work is partially supported by the National Basic Research Program (``973" Program)
of China (Grant Nos 2014CB845800 and 2013CB834900), the National Natural Science Foundation
of China (Grants Nos. 11322328, 11433009, 11543005, 11573014, U1231101, and U1431124), the One-Hundred-Talents Program,
the Youth Innovation Promotion Association (2011231), and the Strategic Priority Research Program
``The Emergence of Cosmological Structures" (Grant No. XDB09000000) of
the Chinese Academy of Sciences, and NASA NNX 13AH50G, 14AF85G and 15AK85G.

\def\apjl{{Astrophys. J. Lett.}}
\def\apj{{Astrophys. J.}}
\def\apjs{{Astrophys. J. Suppl}}
\def\aj{{Astron. J.}}
\def\aap{{Astron. Astrophys.}}
\def\aaps{{Astron. Astrophys. Suppl.}}
\def\apss{{Ap. Sp. Sci.}}
\def\araa{{Ann. Rev. Astron. Astroph.}}
\def\mnras{{Mon. Not. R. Astron. Soc.}}
\def\nat{{Nature}}
\def\science{{Science}}
\def\pasj{Pub. Astron. Soc. Japan}


\end{document}